%% file: main.tex
\documentclass[lettersize,journal]{IEEEtran}
\usepackage{amsmath,amsfonts,amssymb}
\usepackage{algorithmic}
\usepackage{booktabs}
\usepackage{color}
\usepackage{algorithm}
\usepackage{array}
\usepackage[caption=false,font=normalsize,labelfont=sf,textfont=sf]{subfig}
\usepackage{textcomp}
\usepackage{stfloats}
\usepackage{url}
\usepackage{multirow}
\usepackage{verbatim}
\usepackage{graphicx}
\usepackage{cite}
\usepackage{CJKutf8}
\usepackage{soul}
\usepackage{pifont}

\DeclareMathOperator*{\argmax}{argmax}
\ifCLASSINFOpdf
\else
\fi

\begin{document}
\title{Code-switching Speech Recognition Under the Lens: Model- and Data-Centric Perspectives}
\author{Hexin Liu, 
        Haoyang Zhang, 
        Qiquan Zhang, 
        Xiangyu Zhang, 
        Dongyuan Shi,
        Eng Siong Chng, 
        Haizhou Li
\thanks{}}

\markboth{submission to IEEE Transactions on Audio, Speech, and Language Processing}%
{Shell \MakeLowercase{\textit{et al.}}: Bare Demo of IEEEtran.cls for IEEE Journals}

\maketitle
\begin{abstract}
 Code-switching automatic speech recognition (CS-ASR) presents unique challenges due to language confusion \textcolor{black}{that violates the monolingual assumption} and accent bias that blurs the phonetic boundaries. Although the constituent languages may be individually high-resource, the scarcity of annotated code-switching data further compounds these challenges. In this paper, we systematically analyze CS-ASR from both model-centric and data-centric perspectives. By comparing state-of-the-art algorithmic methods, including language-specific processing and auxiliary language-aware multi-task learning, we discuss their varying effectiveness across datasets with different linguistic characteristics. On the data side, we first investigate TTS as a data augmentation method. By varying the textual characteristics and speaker accents, we analyze the impact of language confusion and accent bias on CS-ASR. To further mitigate data scarcity and enhance textual diversity, we propose a prompting strategy by simplifying the equivalence constraint theory (SECT) to guide large language models (LLMs) in generating linguistically valid code-switching text. The proposed SECT outperforms existing methods in ASR performance and linguistic quality assessments, generating code-switching text that more closely resembles real-world code-switching text. When used to generate speech–text pairs via TTS, SECT proves effective in improving CS-ASR performance. Our analysis of both model- and data-centric methods underscores that effective CS-ASR requires strategies to be carefully aligned with the specific linguistic characteristics of the code-switching data.
\end{abstract}


\IEEEpeerreviewmaketitle

\section{Introduction}
\IEEEPARstart{C}{ode-switching} involves the alternation between two or more languages within spontaneous multilingual speech~\cite{sitaram2019survey}. Intra-sentence code-switching occurs when the language changes within a single sentence, while inter-sentential code-switching involves language alternation at sentence boundaries and is commonly treated as multilingual~\cite{pratapa2018language, liuxsa}. Despite significant advancements in monolingual speech processing tasks, such as automatic speech recognition~(ASR) and text-to-speech synthesis~(TTS)~\cite{conformer, survey, ren2019fastspeech, cosyvoice}, code-switching speech processing remains challenging. \textcolor{black}{The challenges arise from} both the complex linguistic characteristics inherent to code-switched speech and the scarcity of annotated data.

\begin{figure}[t]
  \centering
  \includegraphics[width=\linewidth]{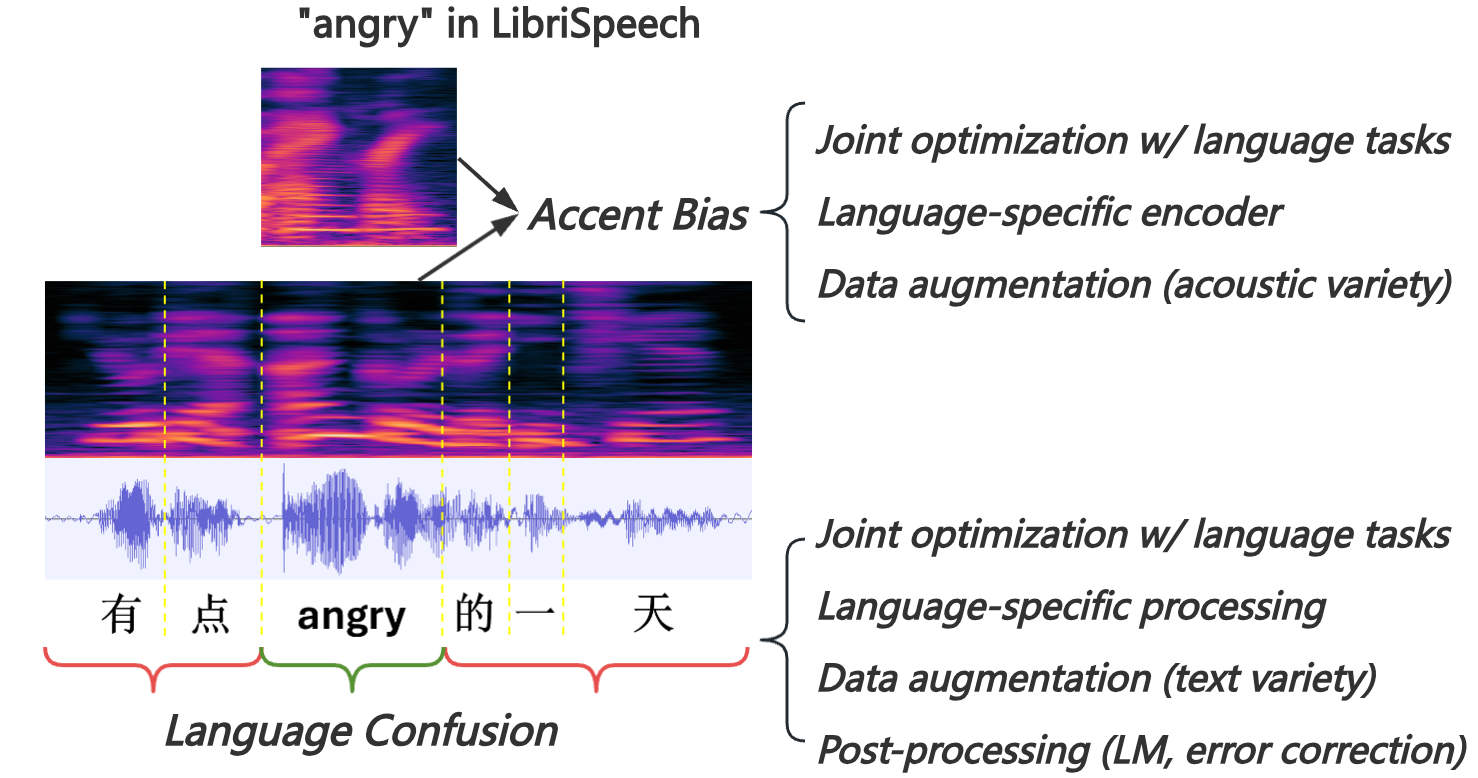}
  \caption{\textcolor{black}{Overview of existing methods in code-switching ASR.}}
  \label{fig:overview}
\vspace{-0.2cm}
\end{figure}

The inherent challenges in code-switching speech processing can be broadly attributed to two primary factors. We define \textcolor{black}{the inherent challenges} as accent bias and language confusion \textcolor{black}{as depicted in Figure~\ref{fig:overview}}. Speakers may exhibit native-like fluency in multiple languages and seamlessly alternate between them during spontaneous conversation. However, accent, particularly common among L2 speakers, can blur the phonetic boundaries between the two languages. We define this phenomenon as accent bias, which complicates both acoustic modeling and the mapping from acoustic features to token embeddings. In \textcolor{black}{parallel}, code-switching introduces language confusion within the matrix language, the dominant or structural language in a code-switching utterance. This confusion is primarily associated with text. \textcolor{black}{It} arises from the insertion of elements from an embedded language that disrupts the expected syntactic and phonological patterns during language modeling~\cite{guzman17_interspeech}. Moreover, in naturally bilingual communities, the matrix and embedded languages may alternate. \textcolor{black}{This alternation leads to} increasingly ambiguous boundaries between \textcolor{black}{the two languages}. This linguistic complexity presents additional challenges. As a result, even large-scale pre-trained multilingual speech models, such as Whisper~\cite{whisper}, exhibit degraded performance on code-switching data. \textcolor{black}{This performance drop occurs despite their high} performance in both the matrix and embedded languages individually. 

Accent bias has typically been mitigated from a data-centric perspective through adaptation with accented speech~\cite{address_accent, yang2023adapting}. Existing studies have attempted to address the language confusion in code-switching speech recognition~(CS-ASR) by integrating language discriminative information. One line of research has sought to factorize the recognition process into separate monolingual components~\cite{bi_encoder, yan2022joint}. To effectively coordinate the language-specific processing, an additional divide-and-conquer strategy \textcolor{black}{is often required. Common choices include} mixture-of-experts (MoE) framework or factorization methods~\cite{bi_encoder, yan2022joint, lae,zhang22x_interspeech, song22e_interspeech, wang2025camel}. Another important stream of work has pursued a unified multi-task paradigm~\cite{zeng19_interspeech, tseng2021mandarin, liu23_icassp, liu2024enhancing, align}. Such approaches involve multi-tasking learning with an auxiliary language identification~(LID) or diarization (LD) module to enrich the unified model with frame- or token-level language information~\cite{tseng2021mandarin, li2013spoken, liuxsa}. \textcolor{black}{In addition, post-processing techniques have proven effective in mitigating language confusion. These include late fusion with a language model (LM) and error correction~\cite{ger, align}.}

Prior research has explored data augmentation strategies to mitigate data scarcity for low-resource and code-switching ASR~\cite{baas22_interspeech, yang2025enhancing, ogun2025exhaustive}. Traditional data augmentation methods for ASR include speed perturbation and SpecAugment~\cite{speed_pert, specaug}. These approaches expand training data by altering the acoustic properties of the original speech signal, such as modifying the speaking rate or applying spectrogram masking. Voice conversion (VC) and TTS provide another solution by introducing speaker variability or generating speech signals from text data. Pre-trained VC and TTS systems typically underperform for low-resource languages due to limited tokenizer exposure and insufficient linguistic coverage~\cite{cosyvoice}. In contrast, code-switching scenarios involving high-resource language pairs, such as English and Mandarin, moderately circumvent this limitation, as both languages are well-represented in large-scale TTS training corpora~\cite{zhang2024speaking, wang2025maskgct}. Consequently, synthesizing code-switching speech using TTS and VC systems pre-trained on high-resource constituent languages offers a practical and effective data augmentation strategy to improve the CS-ASR performance ~\cite{cs_fleurs}. Existing research has also explored synthesizing code-switching text to increase textual diversity for TTS-based data augmentation in ASR training~\cite{chang19_interspeech, 10096317, 10389644}. Earlier approaches typically relied on rule-based frameworks or generative models such as generative adversarial networks (GANs)~\cite{10096317, chang19_interspeech}. More recently, large language models (LLMs) have been adopted for this task~\cite{kuwanto2024linguistics, 10389644, cs_fleurs}. \textcolor{black}{They} have proven effective in generating code-switching text data that more aligns with real-world code-switching sentences. 


\textcolor{black}{In this paper, we investigate CS-ASR from both model- and data-centric perspectives. The contributions of this work are summarized as follows:}
\begin{itemize}
    \item \textcolor{black}{\textbf{Model-centric analysis.} We systematically review and compare recent algorithmic approaches, including language-specific processing and multi-task learning with language-aware tasks. While our analysis reveals the effectiveness of algorithmic methods, their performance is often limited by accent bias. This observation motivates a data-centric investigation into the respective impacts of accent bias and language confusion.}
    \item  \textcolor{black}{\textbf{Data-centric analysis.} We explore TTS-based speech synthesis and LLM-based text generation to address data scarcity. By disentangling the effects of language confusion and accent bias, our findings demonstrate that syntactic and lexical characteristics exert a more significant influence on CS-ASR performance than speaker accents.} 
    \item \textcolor{black}{We introduce a simplified Equivalence Constraint Theory (SECT) prompting strategy that bridges formal linguistic theory with LLM capabilities. This method outperforms existing prompting methods in linguistic quality and downstream ASR performance.}
    \item \textcolor{black}{By comparing model- and data-centric approaches, we summarize practical strategies for different code-switching scenarios, offering insight for advancing CS-ASR research and applications.}
\end{itemize}

\section{Model-centric methods}
\subsection{Unified model and auxiliary tasks}
Existing works have demonstrated that CS-ASR can be achieved via a general ASR system with a unified vocabulary comprising all multilingual tokens~\cite{espnet}. Let the input speech signal be represented as a sequence of acoustic feature vectors $\mathbf{X}=(\mathbf{x}_{t} \in \mathbb{R}^{F}\mid t=1, \ldots, T)$, where $F$ is the feature dimension and $T$ is the sequence length. The corresponding tokenized text sequence is defined as $Y=(y_n \in \mathcal{V} \mid n=1, \ldots, N)$, where $\mathcal{V} = \mathcal{V}^{(L_1)} \cup \mathcal{V}^{(L_2)}$ is the vocabulary comprising vocabularies of languages $L_1$ and $L_2$, and $N$ is the token sequence length. The speech recognition process is achieved by computing the conditional probability distribution of the token sequence given the acoustic features
\begin{equation}
  p_{\mathrm{asr}}\left(Y|\mathbf{X}\right) \approx \prod_{n=1}^{N} p\left (y_{n}|y_{1:n-1},\mathbf{X}\right),
\label{eq:att_decode}
\end{equation}
where each token probability is conditioned on the acoustic features and the previously generated tokens. This formulation corresponds to the attention-based sequence-to-sequence model, in which decoding is performed autoregressively. The predicted sequence $\widehat{Y}$ is obtained by maximizing the above conditional probability.
\begin{equation}
    \widehat{Y} = \argmax_Y \big \{ \mathrm{log}p_{\mathrm{asr}}\left(Y|\mathbf{X}\right)\big\}.
  \label{eq:decoding}
\end{equation}

\begin{figure}[t]
  \centering
  \includegraphics[width=\linewidth]{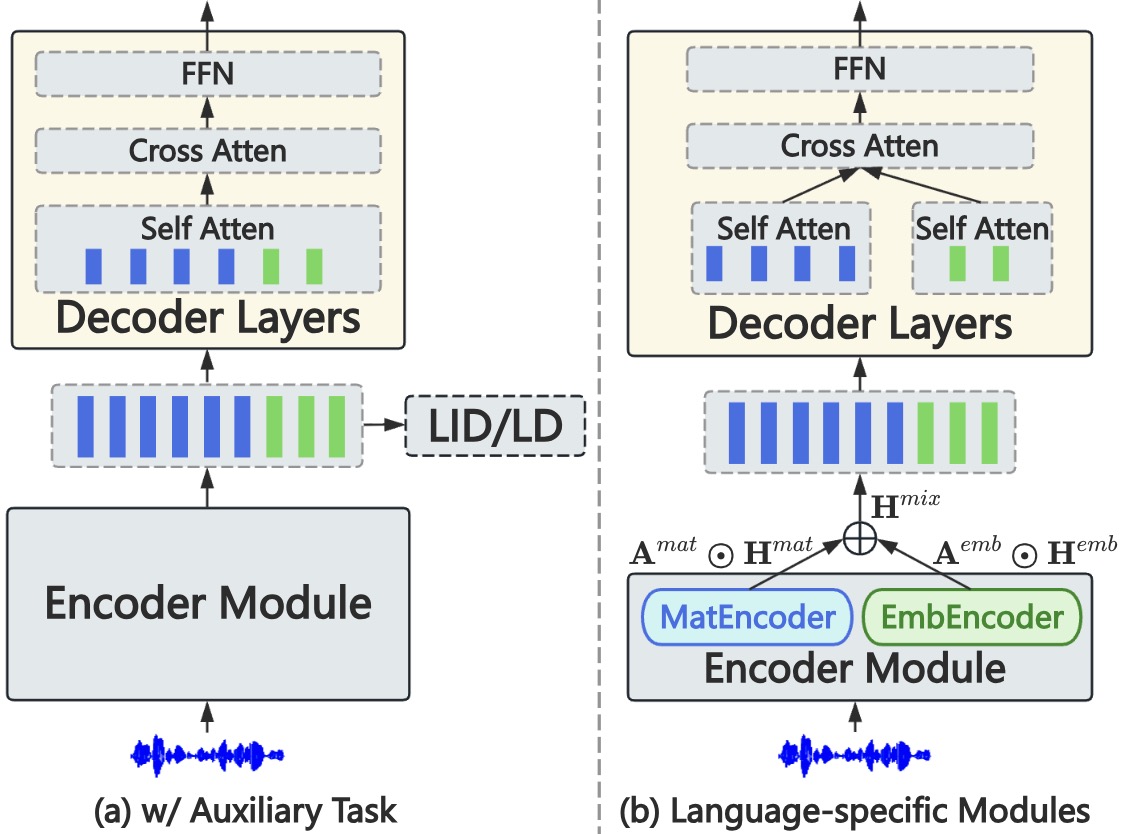}
  \caption{Improved CS-ASR algorithms with (a) multi-task learning with auxiliary task or (b) language-specific modules, where blue and green colors denote two languages, long and short units denote speech and text representations, respectively.}
  \label{fig:models}
\end{figure}

However, the above approach rarely makes explicit use of language information. To mitigate this limitation, recent studies have explored enriching the model via multi-task learning with auxiliary language-aware tasks. \textcolor{black}{As shown in Figure~\ref{fig:models}(a), these tasks are closely related to language information, such as} LD or LID~\cite{zeng19_interspeech, tseng2021mandarin, liu23_icassp}. These tasks are typically performed at the frame or token level to achieve fine-grained language change detection. The CS-ASR model is then optimized with a multi-task objective, formulated as the weighted sum of the ASR loss and the auxiliary language-aware loss as
\begin{equation}
  \mathcal{L}_{\mathrm{multi}}=\alpha \mathcal{L}_{\mathrm{asr}} + \left ( 1-\alpha  \right )\mathcal{L}_{\mathrm{lang}},
  \label{eq:loss_joint}
\end{equation}
where $\mathcal{L}_{\mathrm{asr}}$ denotes the ASR loss, $\mathcal{L}_{\mathrm{lang}}$ represents the loss from the auxiliary language-aware task, and $\alpha$ is a multi-task learning parameter.

\subsection{Language-specific Modules} 
\label{sec:language_spe_module}
Another important line of research focuses on incorporating language-specific processing within CS-ASR models\textcolor{black}{, as presented in Figure~\ref{fig:models}(b)}. Language-specific encoders process the acoustic feature vectors $\mathbf{X}$ separately into hidden representations $\mathbf{H}=(\mathbf{h}_{t} \in \mathbb{R}^{D}\mid t=1, \ldots, T^{\prime})$ before performing a weighted sum of them. The weights $\mathbf{A}^{mat/emb}=(a_{t} \in [0,1] \mid t=1, \ldots, T)$ are frame-level language posteriors computed by a built-in LID module. These are computed via
\begin{align}
\mathbf{H}^{mat} &=\mathrm{MatEncoder}\left( \mathbf{X} \right), \label{eq:bi_encoder1} \\
\mathbf{H}^{emb} &= \mathrm{EmbEncoder}\left( \mathbf{X} \right), \label{eq:bi_encoder2} \\
\mathbf{H}^{mix} &= \mathbf{A}^{mat} \odot \mathbf{H}^{mat} + \mathbf{A}^{emb} \odot\mathbf{H}^{emb}, \label{eq:bi_encoder3}
\end{align}
where $mat$ and $emb$ denote the matrix and embedded language, respectively, and $\odot$ is element-wise product. We use $\mathrm{MatEncoder}\left ( \cdot \right )$ and $\mathrm{EmbEncoder}\left ( \cdot \right )$ to represent the computations within language-specific encoders. The mixed hidden representations are denoted as $\mathbf{H}^{mix}$ and are fed into the decoder layers. In this manner, the system approximates monolingual ASR at the frame level to reduce language confusion. Ideally, when the input is monolingual, the corresponding encoder should receive a weight close to one, while the other encoder is assigned a weight close to zero. This mechanism ensures that each $\mathbf{h}_{t}^{mix}$ corresponding to one language closely resembles the output of the corresponding monolingual encoder, which suppresses interference from the non-target language.

Language-specific processing within the ASR decoder typically involves language-specific self-attention computations. This is achieved by using separate token embedding layers for two languages. We define $\mathbf{Y}^{mix}=(\mathbf{y}_{\_,n}^{mat/emb} \in \mathbb{R}^{D}\mid n=1, \ldots, N)$ as the token embedding sequence comprising language-specific token embeddings. The language-specific token embeddings $\mathbf{Y}^{mat}=(\mathbf{y}_{n',n}^{mat} \in \mathbb{R}^{D}\mid n'=1, \ldots, N^{mat})$ and $\mathbf{Y}^{emb}=(\mathbf{y}_{n'',n}^{emb} \in \mathbb{R}^{D}\mid n''=1, \ldots, N^{emb})$ are fed into their respective self-attention module. Here, we use $n'$ and $n''$ to denote the indices within $\mathbf{Y}^{mat}$ and $\mathbf{Y}^{emb}$, respectively. The above can be illustrated as
\begin{align}
N&=N^{mat}+N^{emb},\\
\mathbf{Y}_{o}^{mat}&=\mathrm{SelfAtten}^{mat}(\mathbf{Y}^{mat}),\\
\mathbf{Y}_{o}^{emb}&=\mathrm{SelfAtten}^{emb}(\mathbf{Y}^{emb}),
\label{eq:bi_decoder}
\end{align}
where $\mathrm{SelfAtten}^{mat}(\cdot)$ and $\mathrm{SelfAtten}^{emb}(\cdot)$ denote the computations within the self-attention module for matrix and embedded languages. The embeddings $\mathbf{Y}_{o}^{mat}$ and $\mathbf{Y}_{o}^{emb}$ represent the outputs of the respective self-attention modules. They are subsequently combined back into a unified $N$-length embedding sequence $\mathbf{Y}_{o}^{mix} = (\mathbf{y}_{\_,n,o}^{mat/emb} \in \mathbb{R}^{D} \mid n = 1, \ldots, N)$ before the cross-attention computations.

The auxiliary language-aware task is typically assigned a relatively small weight (i.e., high $\alpha$ value) compared to the ASR task during training, indicating its minor role in guiding the optimization. By contrast, language-specific modules contribute symmetrically across languages. Each module processes features in the same manner~\cite{bi_encoder, song22e_interspeech}, regardless of the relative proportions of matrix and embedded languages in the training data. In the training stage, language-specific modules may require pre-training on \textcolor{black}{constituent languages} to learn language-discriminative information before being applied to code-switching data, whereas multi-task learning can be performed solely on code-switching data~\cite{tseng2021mandarin, liu23_icassp, align}. During inference, the auxiliary language branch is typically inactive, the decoding process is thus identical to that of a unified model. In contrast, language-specific modules require two passes through the encoder or decoder, introducing additional computational overhead and higher decoding latency. For example, the bi-encoder method exhibits real-time factors (RTF) and latency more than twice those of a model with a single encoder. These differences highlight that auxiliary tasks can enhance CS-ASR with minimal overhead, while language-specific modules deliver more explicit language discrimination at the expense of efficiency.

\section{Data-centric analysis}
\subsection{Data Scarcity and Monolingual Utilization}
Code-switching, particularly intra-sentential code-switching, occurs far less frequently than monolingual speech. Moreover, annotating code-switching data requires bilingual expertise, which further limits the availability of resources. As a result, code-switching ASR suffers from data scarcity, even though constituents can be individually high-resource languages. 

To address this challenge, existing research has leveraged resources from the constituent languages and explored methods for synthesizing code-switching text and speech. Several studies have shown that monolingual data, particularly from the matrix language, can support the development of CS-ASR systems~\cite{bi_encoder, yan2022joint, zs_tts_msc, zs_brian_icassp23, zs_knn, wang2025camel}. For example, \cite{zs_tts_msc, zs_brian_icassp23, zs_knn} investigated building CS-ASR models using only monolingual data. These approaches typically employ language-specific connectionist temporal classification (CTC)~\cite{ctc, zs_brian_icassp23}. \textcolor{black}{Here,} decoding is first performed independently for each language and then combined through joint decoding. Tokens in the matrix-language sequence are \textcolor{black}{then} substituted with the corresponding embedded-language tokens when appropriate. However, these works also indicate that subsequent fine-tuning on code-switching data provides substantial performance gains. Moreover, an excessive reliance on monolingual data can result in training imbalance and degrade performance on code-switching speech. Therefore, augmenting code-switching data is crucial for effectively addressing this scarcity.
\subsection{Code-switching speech synthesis}
\begin{figure}[t]
  \centering
  \includegraphics[width=\linewidth]{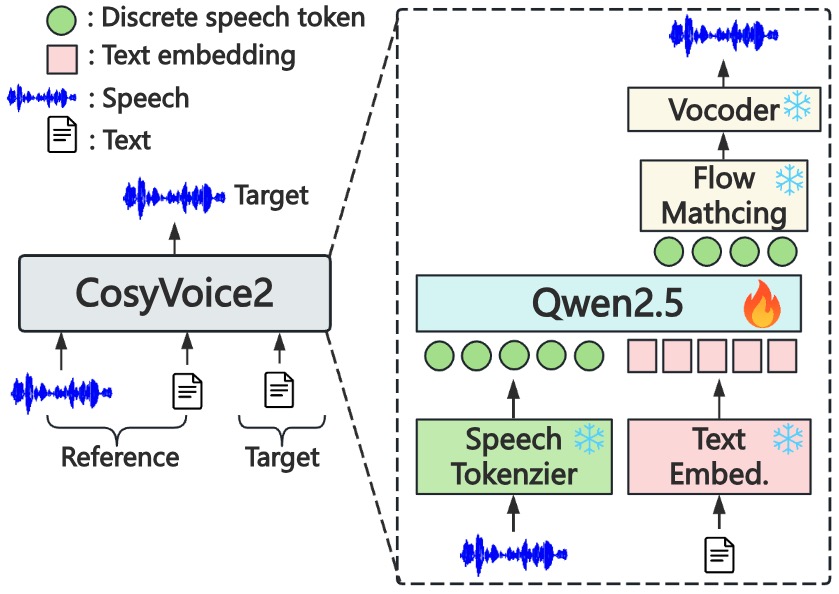}
  \caption{The CosyVoice 2 model used for TTS, only the LLM module within CosyVoice 2 is updated during fine-tuning.}
  \label{fig:CosyVoice2}
\end{figure}
Recent advances in TTS models have enabled the use of synthesized speech for augmenting ASR training~\cite{yang2025enhancing}, particularly mitigating the data scarcity in low-resource scenarios. In this work, we investigate the effectiveness of TTS-based data augmentation for improving CS-ASR with a CosyVoice 2 model~\cite{cosyvoice}, as shown in Figure~\ref{fig:CosyVoice2}. CosyVoice 2 is an LLM-based TTS method, comprising a speech tokenizer, an LLM, and a conditional flow matching (CFM) module~\cite{lipman2023flow}. The speech tokenizer is developed by integrating finite scalar quantization into the encoder module of an ASR model~\cite{mentzer2024finite, an2024funaudiollm}, converting continuous speech signals into discrete speech tokens. The LLM then serves as a text–speech language model, generating speech tokens autoregressively with the input text tokens. Finally, the conditional flow matching (CFM) module decodes the speech tokens into a Mel spectrogram, conditioned on the reference speech, the speaker embeddings, and the LLM-generated tokens from the target text.

In this work, we adopt LLM-based TTS rather than traditional TTS for code-switching speech synthesis. This is because the LLM is pre-trained with machine translation tasks and thus inherently exhibits cross-lingual capability, which is advantageous in code-switching scenarios. In addition, the model is pretrained on massive English and Mandarin data. We only update the LLM module within CosyVoice 2 during fine-tuning on the target code-switching data since the speech tokenizer and flow matching modules are not open-sourced. 

To analyze the effects of text style and accent bias on CS-ASR, we simulate diverse code-switching speech–text pairs via TTS. The synthetic speech is generated using varying combinations of target and reference text and speech domains\textcolor{black}{. It thus} inherits the acoustic characteristics and speaker identity from the reference (i.e., prompt) speech while adhering to the textual pattern of the target text. This design allows us to isolate and evaluate the individual and combined effects of stylistic and accentual variation on ASR robustness.

\subsection{Code-switching text generation}
\label{sec:cs_text}
CS-ASR is inherently a cross-lingual task, as it involves dynamic alternation between two or more languages within a single utterance or conversation. A natural solution to this challenge is to develop a multilingual ASR model equipped with text translation capabilities across language pairs. However, such an approach typically demands large-scale supervised multilingual and parallel data. However, such data are often unavailable, particularly for spontaneous speech commonly observed in code-switching scenarios. Consequently, the applicability of translation-augmented multilingual ASR models remains limited in real-world settings due to severe data scarcity.

An alternative strategy focuses on directly optimizing ASR models with code-switching data. However, this approach is often limited by data scarcity. Although TTS can alleviate the shortage of speech resources, generating diverse and natural code-switching text remains a major challenge. This difficulty arises from the wide variability in code-switching patterns\textcolor{black}{. These patterns may} involve arbitrary combinations of syntactic structures, lexical substitutions, and language change points. Consequently, synthesizing high-quality code-switching text is crucial, as it forms the basis for constructing paired speech–text data via TTS~\cite{10389644, cs_fleurs}.

\begin{figure}[t]
  \centering
  \includegraphics[width=\linewidth]{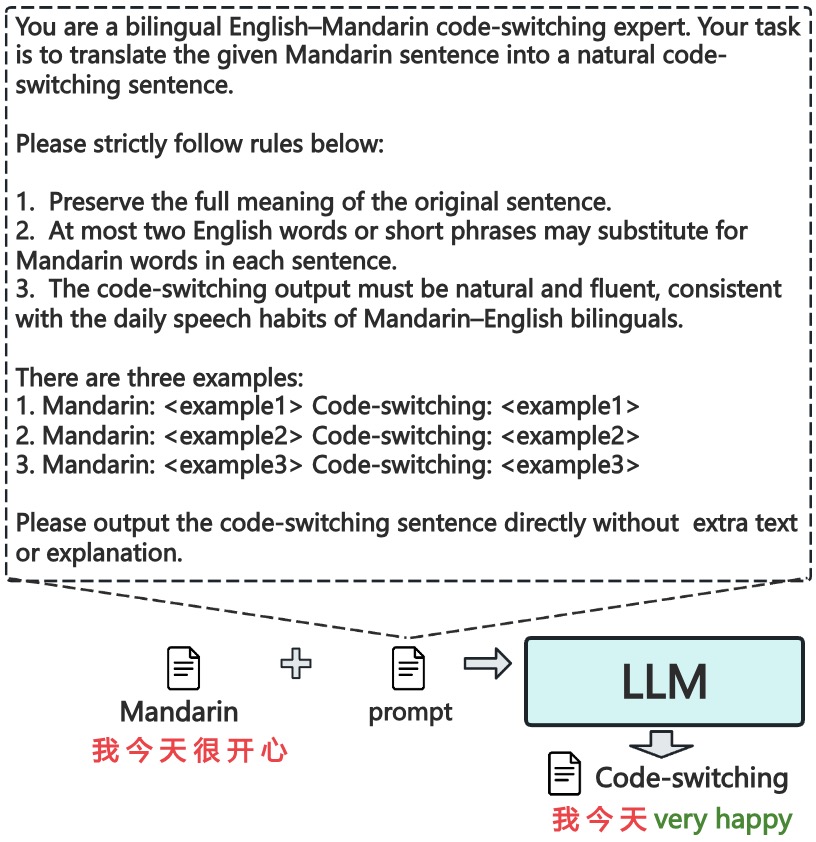}
  \caption{An example of translating Mandarin text to English-Mandarin Code-switching text via SECT-guided LLM, the proposed SECT-based prompt is shown in the dashed block.}
  \label{fig:llm}
\end{figure}

LLMs present a promising solution for this task due to their inherent multilingual and translation capabilities. However, unconstrained generation may lead to unnatural or linguistically implausible switches~\cite{10389644}. To address this, linguistic constraints such as the ECT can be incorporated to guide generation~\cite{ect, pratapa-etal-2018-language, chang19_interspeech,kuwanto2024linguistics, gcm}. ECT constrains code-switching to occur only at points where the grammatical structures of both languages align. When implemented in prompt-based generation, ECT can be simplified since LLMs already internalize linguistic regularities to maintain syntactic compatibility and naturalness. 

Building on this observation, we propose to integrate a simplified version of ECT (SECT) into the prompt design for LLM-based code-switching text generation. Beyond leveraging the inherent syntactic knowledge of LLM, SECT offers improved computational efficiency and avoids imposing overly rigid constraints compared to ECT.

As depicted in Figure~\ref{fig:llm}, monolingual sentences are translated from a high-resource matrix language (Mandarin) into their English-Mandarin code-switching counterparts via the prompt-guided LLM. This process is governed by SECT, which is formulated as three rules. The first rule specifies the matrix language and prevents the LLM from making substantial modifications to the original sentence. The second rule introduces a constraint to control the degree of code mixing. \textcolor{black}{This constraint }can be adjusted depending on the desired characteristics of the target code-switching text. Finally, the third rule ensures that the generated code-switching sentences adhere to the syntactic structure of the matrix language while embedding content words from the embedded language at linguistically appropriate switch points. To further guide the model, we adopt an in-context learning strategy by providing three illustrative examples within the prompt. Each example consists of an input and output, being a monolingual sentence and its code-switching counterpart. 

Compared with existing LLM-based approaches for code-switching text generation~\cite{kuwanto2024linguistics, cs_fleurs}, SECT offers a more convenient solution\textcolor{black}{. This is achieved by} exploiting the LLM’s intrinsic knowledge and eliminating the need for feeding parallel matrix–embedded language text into each prompt. By leveraging the generative capability of LLMs under SECT constraints, the proposed method synthesizes high-quality code-switching text directly from monolingual data\textcolor{black}{. This generated text} can then be used for downstream ASR training.

\section{Datasets and experiments}
\subsection{Datasets}
\label{sec:datasets}

We evaluated CS-ASR performance using two widely studied English–Mandarin code-switching speech corpora: the ASRU 2019 Code-Switching Challenge dataset and the SEAME dataset~\cite{seame, shi2020asru}. The ASRU data comprises four subsets: a Mandarin-only training set, an intra-sentence code-switching training set, two intra-sentence code-switching development sets (the dev1 set is used), and an intra-sentence code-switching test set. 

The SEAME dataset is a corpus of spontaneous conversational speech recorded in Singapore and Malaysia, featuring monolingual and code-switching utterances~\cite{seame}. Following the partitioning protocol described in~\cite{zeng19_interspeech}, we divided the dataset into a 101.5-hour training set and two test sets, denoted as $\text{dev}_{\texttt{man}}$ and $\text{dev}_{\texttt{sge}}$. A 4.9-hour validation set was further extracted from the training data to monitor training progress. 

The LibriSpeech dataset and the Mandarin-only training set from the ASRU 2019 Code-Switching Challenge are also used to evaluate monolingual ASR performance and to pre-train multilingual models from scratch. To ensure balanced training, only the train-clean-100 and train-clean-360 dataset of LibriSpeech are used during multilingual pre-training.

\input{Tables/dataset}

The two code-switching datasets differ primarily in accent and syntactic structure. Since the ASRU dataset is recorded in mainland China, data within this corpus is characterized by a Chinese accent and contains Mandarin-centric text, with a grammatical structure primarily framed in Mandarin and embedded English segments. Sentences within the ASRU training and development sets, on average, contain 1.6 English words and 8.6 Chinese characters. As opposed to the ASRU data, the SEAME dataset is recorded in Singapore and Malaysia and features speakers with South-East Asian accents. We also compute the code mixing index~(CMI) as defined in~\cite{gamback-das-2016-comparing}, to measure the degree of language mixing of code-switching text data. The CMI values of the ASRU and SEAME training sets are 17.0 and 13.6, respectively, while the CMI of SEAME \textcolor{black}{training data} increases to 24.2 when monolingual utterances are excluded. Therefore, SEAME data contains more frequent code-switching than the ASRU dataset, largely due to the bilingual education systems and language policies in these regions~\cite{dixon2005bilingual}. The higher frequency and fluidity of language switching suggest that SEAME poses greater challenges for code-switching ASR, particularly as the matrix and embedded languages can alternate across sentences~\cite{auer1999codeswitching, syntactic_LM}.

\input{Tables/language_ratio}
Detailed statistics for all datasets are provided in Table~\ref{tab:data}, and Table~\ref{tab:seame} reports the duration ratios of each language based on utterance-level annotations for code-switching test sets. CMI values of the ASRU and SEAME training data are 17.0 and 13.6, respectively, where that of SEAME training data is 24.2 after excluding monolingual utterances.
\subsection{Experiment setup and model configuration}
Experiments with models trained from scratch, such as Conformer, were conducted using ESPNet~\cite{espnet}. For training on the SEAME and ASRU datasets, we adopted the hyperparameter settings from~\cite{liu23_icassp} for Conformer and from~\cite{mamba_in_speech} for Branchformer and ConExtBiMamba. Experiments involving pre-training on monolingual datasets followed the hyperparameters specified in the LibriSpeech recipe provided by ESPNet. The resulting model is referred to as Conformer-L Mono. Conformer-L has a larger hidden dimension of 512, in contrast to the 256-dimensional hidden size used in the Conformer models. This setup is also employed for Conformer-L Mix, which was \textcolor{black}{developed} on a mixture of code-switching and monolingual datasets. Inference was performed using the average of the weights from the ten best-performing models on the development sets, with a beam size of five.

Experiments with Whisper models were conducted using Whisper-small~\cite{whisper}. Fine-tuning employed the AdamW optimizer, with the learning rate linearly warmed up from 0 to $1 \times 10^{-5}$ over 10,000 update steps, followed by a linear decay schedule. Training used a batch size of eight with gradient accumulation over two steps. For inference, we adopted greedy decoding with model weights averaged from the three best-performing checkpoints on the development set. Language prompts $<$zh$>$ and $<$en$>$ were applied to the ASRU and SEAME datasets, respectively. LoRA with a rank of 24 was integrated into the feed-forward networks of Whisper~\cite{hu2022lora}. Following the Bi-encoder mechanism~\cite{bi_encoder}, which is described in equations~(\ref{eq:bi_encoder1})–(\ref{eq:bi_encoder3}), Bi-LoRA was pre-trained separately on LibriSpeech and ASRU Mandarin, and subsequently fine-tuned on either ASRU or SEAME, with each LoRA configured to a rank of 12. \textcolor{black}{This ensures a fair comparison between general LoRA and bi-LoRA adaptation methods while achieving computational efficiency.}

Speech synthesis was performed using the open-source TTS model CosyVoice 2~\cite{cosyvoice}, which was fine-tuned on the target domain data before synthesis. \textcolor{black}{ CosyVoice 2 is pre-trained on massive English and Mandarin data, providing rich prior bilingual knowledge. This is critical for English-Mandarin code-switching, as the model has already learned the joint distribution and phonetic transitions between these two languages. This prior knowledge ensures that the synthesized code-switching data is phonetically accurate, providing higher effectiveness for developing CS-ASR models.} Speech samples from different datasets as prompt speech were used to simulate various accents: LibriSpeech for American English, ASRU for Mandarin spoken in mainland China, and SEAME for Southeast Asian-accented English and Mandarin. 

Code-switching text generation was performed via Qwen 1.5-32B-chat~\cite{qwen}, and Deepseek-R1 API~\cite{liu2024deepseek}, respectively. As shown in Figure~\ref{fig:llm}, we employed a prompt containing several examples of converting monolingual sentences into their code-switching counterparts. The LLMs then generated code-switching sentences by referencing the prompt and applying the transformation to the target monolingual inputs. 
\subsection{Performance measurement}
We evaluated the CS-ASR models using the mixed error rate (MER), which combines word error rate (WER) for English and character error rate (CER) for Mandarin. English and Mandarin ASR performance was reported in WER and CER, respectively.

In addition to the ASR performance, we adopted an LLM-as-a-judge method to evaluate the quality of texts generated by LLMs. Specifically, GPT-4o assigns a score from 1 to 10 to each sample with reference to the criteria described in Table~\ref{tab:prompt_comparison}. The final score is computed as the average across all text samples. \textcolor{black}{Human evaluation was conducted involving ten bilingual graduate students proficient in both English and Mandarin. The evaluators were instructed to assess the generated text samples for each method using the same scoring criteria as the LLM-as-a-judge framework.} We also evaluate the similarity in language confusion between real text and LLM-generated text by comparing their CMI \textcolor{black}{and perplexity (PPL). The PPL is calculated with a Transformer-based language model (LM) developed on the training set of ASRU data via ESPNet. The PPL of the test data serves as the ground-truth baseline.}
\section{Results and Analysis}
\subsection{Comparison of model-centric methods}
We conduct a comparison of existing model-centric approaches to provide insight into how different algorithmic strategies are designed for CS-ASR. Table~\ref{tab:sota_compare} presents a comparison between models trained from scratch and those fine-tuned on Whisper-small models. Table~\ref{tab:pre-train} illustrates the impact of pre-training and fine-tuning strategies on CS-ASR performance.

Results in Table~\ref{tab:sota_compare} suggest that algorithmic strategies, such as adopting more advanced model architectures or incorporating language-aware designs, yield significant performance gains over the baseline on the ASRU dataset. In contrast, these strategies lead to only moderate improvements on the SEAME dataset. These can be attributed to the characteristics of the two corpora. Compared to SEAME, the ASRU dataset contains fewer language switches and thus exhibits a lower degree of language confusion, such as less frequent code-switching and discourse markers. As a result, its syntactic and lexical characteristics are more closely aligned with the matrix language, resulting in less language confusion. In addition to more complex textual patterns, SEAME data is characterized by Southeast Asian accents. The above makes the two languages less distinguishable and introduces greater challenges for code-switching ASR models. Consequently, language-specific processing or joint optimization with language-aware tasks provides limited improvement on SEAME.

\textcolor{black}{We further evaluated self-supervised learning (SSL) speech representations by training a Conformer-based ASR system. Results indicate that SSL speech representations significantly improve CS-ASR performance, which aligns with current research~\cite{ssl_review}. Specifically, multilingual features extracted from XLSR outperformed monolingual features extracted from WavLM. This suggests that multilingual pre-training provides more robust cross-lingual information that is beneficial for modeling code-switched speech.}

It is useful to note that CAMEL, which is built upon E-Branchformer, achieves the highest performance on SEAME and ASRU test sets among all models trained from scratch by integrating both language-specific modules and multi-task learning with LD. Instead of directly enriching the encoder with language-discriminative information via multi-task learning~\cite{liu23_icassp}, CAMEL incorporates language information within the ASR decoder. Specifically, the key and values matrices within the LD decoder are extracted before being used to perform cross-attention with the query matrix associated with the token embedding sequence. This suggests that both strategies are beneficial to CS-ASR and highlights the potential of combining them in a complementary and well-coordinated manner to further enhance performance.

As discussed in Section~\ref{sec:language_spe_module}, the language-specific process aims to achieve monolingual ASR due to the limited multilingual capacity of the ASR model. Nevertheless, for large-scale multilingual pre-trained models such as Whisper, the necessity of language-specific processing is substantially diminished. This is supported by the results of comparing adaptation strategies for Whisper-small using LoRA and Bi-LoRA. Specifically, although the bi-LoRA method involves separate pre-training of LoRA modules on the respective languages, it fails to consistently yield higher performance compared to a single LoRA on the SEAME and ASRU test sets. Compared to the integration of language-specific modules, joint optimization with auxiliary tasks, such as LAL and FA-LID, proves more effective in improving CS-ASR performance for large-scale pre-trained multilingual models. However, the use of language-specific modules remains a more flexible and modular solution, as it can be deployed as adapters without updating the base model. In contrast, auxiliary-task-based methods often require updating model parameters, making them relatively heavier for practical deployment, although the auxiliary branch typically introduces only a small increase in model parameters.
\input{Tables/overall}
\subsection{Impact of Pre-training}
\input{Tables/pretrain_2}
\textcolor{black}{When pre-trained models like Whisper are fine-tuned exclusively on code-switching data, the system degrades their performance on existing languages. This is a well-documented phenomenon in the literature~\cite{two_stage_lora}. It is often attributed to catastrophic forgetting or a distributional shift as the model adapts to achieve high performance on an unseen domain.}

\textcolor{black}{To explore solutions to the distributional shift, we also evaluated} the impact of pre-training strategies on CS-ASR performance and presented the results in Table~\ref{tab:pre-train}. Training a Conformer-L model directly on the mix of monolingual data and code-switching data obtains higher performance than sequentially pre-training on monolingual data and fine-tuning on code-switching data. This suggests that CS-ASR systems benefit from both cross-lingual and monolingual knowledge during training. While incorporating code-switching data in the training data moderately degrades monolingual ASR performance, it results in higher overall performance than both the pre-training and the fine-tuning strategies. The pre-trained Whisper-small model demonstrates an inherent ability to handle code-switching due to its large-scale multilingual pre-training. Fine-tuning the Whisper-small model on the ASRU leads to higher performance on both in-domain ASRU Mandarin and code-switching test data, while significantly degrading the performance on the LibriSpeech test-clean set. 

One interesting observation about the performance of Whisper models in CS-ASR tasks is the gap between the ASRU and SEAME datasets. The Whisper-small model reaches a Mixed Error Rate (MER) of 24.9\% on the ASRU test set but performs significantly worse on SEAME, with MERs above 60\% on SEAME test sets. This may be attributed to the translation ability of Whisper models from Mandarin to English. Since Mandarin consistently serves as the matrix language in ASRU and English as the embedded one, this translation ability may help the model recognize the code-switched segments more accurately.

In contrast, the SEAME dataset presents more complex linguistic patterns, where the matrix and embedded languages alternate dynamically across utterances. The absence of English-to-Mandarin translation ability in Whisper may limit its ability to generalize to the code-switching scenarios in SEAME data, especially when English serves as the matrix language. This hypothesis aligns with the statement in Section~\ref{sec:cs_text}, suggesting that the translation capacity, especially in the direction matching the matrix-to-embedded language, can assist in improving CS-ASR performance.

\input{Tables/syn_augment}
\subsection{TTS with original text and speaker variety}
\label{sec:exp_tts_aug}
In this section, we evaluate the impact of TTS-based data augmentation on CS-ASR performance for both the SEAME and ASRU datasets, using only original text data as the target text for TTS synthesis. The corresponding results are presented in Table~\ref{tab:syn_data_aug}. To introduce speaker variation beyond what is seen by the original text data, each target text was paired with a prompt audio randomly sampled from a different speaker within the same dataset. This strategy ensures that the prompt and target did not originate from the same speaker.

While fine-tuning a CS-ASR model on synthesized data outperforms the zero-shot setup, it yields moderately lower performance compared to fine-tuning on real speech. Additionally, fine-tuning the TTS model on target code-switching data prior to speech synthesis provides substantial benefits for downstream ASR performance. 

Results indicate that augmenting training with synthesized speech and original text improves performance on SEAME but degrades performance on ASRU. When augmented with synthesized data three times the size of the original training set, the fine-tuned model achieves the best WER of 12.7\% and 17.7\% on the two SEAME test sets, respectively. However, this data augmentation strategy leads to performance degradation for ASRU. This discrepancy can be attributed to differences in textual complexity between the SEAME and ASRU datasets. \textcolor{black}{We also compare this augmentation strategy with the traditional speed perturbation method. The results show that augmenting data with synthesized speech and original text yields higher performance than speed perturbation under the same data scale. Moreover, combining both augmentation methods leads to further performance gains on the SEAME dataset.}

The above is consistent with our finding in the previous section that the ASRU data exhibits less language confusion compared to the SEAME data. Therefore, a model fine-tuned on ASRU data is able to learn its structural patterns effectively from the original training data, and additional synthesized ASRU data introduces redundancy or noise rather than beneficial linguistic diversity.

Furthermore, the results in Tables~\ref{tab:sota_compare} and \ref{tab:syn_data_aug} indicate that model-centric approaches are more beneficial for code-switching data with lower language confusion, where textual patterns are more easily learned.
\input{Tables/tts_data_mismatch}
\subsection{Impact of accents and language confusion on ASR}
Since the $\text{dev}_{\texttt{man}}$ set has a CMI closer to the SEAME training data, the performance gap between $\text{dev}_{\texttt{man}}$ and $\text{dev}_{\texttt{sge}}$ presented in Table~\ref{tab:sota_compare} suggests that the model performs worse on data with a higher degree of code mixing. To further examine how accent and language confusion in code-switching data affect ASR performance, we synthesized speech using reference prompts and target texts from multiple datasets, respectively. The resulting synthetic data were then used to fine-tune the Whisper-small model, and performance was compared across conditions \textcolor{black}{as shown in Table~\ref{tab:mismatch}}.

As a baseline, we fine-tuned the model using in-domain synthesized speech, where the prompts, target texts, and TTS model all originate from the same dataset. As expected, pairing text with a prompt from the same domain yields the highest ASR performance when using synthesized speech for fine-tuning.

To simulate accent mismatch, we paired text samples from the target dataset with prompt speech samples from a different dataset while keeping the TTS model consistent with the target dataset. For example, using a TTS model fine-tuned on SEAME data to synthesize speech for SEAME text with LibriSpeech or ASRU reference results in an accent mismatch. Conversely, the textual mismatch, primarily associated with syntactic and lexical characteristics, was simulated by pairing out-of-domain text with an in-domain prompt and TTS model. 

A key insight from our study is that the textual characteristics of code-switching data has a significantly greater impact on CS-ASR performance than accent variation. Specifically, fine-tuning the CS-ASR model on data with textual mismatch results in substantially greater performance degradation compared to fine-tuning on data with accent mismatch. This performance gap can be attributed to differences in textual complexity. The SEAME dataset contains more fluent and frequent language switching within utterances, leading to greater language confusion than the structurally simpler ASRU dataset. When SEAME-style TTS model and prompt, which encourage frequent language alternations, are paired with ASRU text, the mismatch causes the model to learn inconsistent switching patterns, ultimately impairing generalization. It is also worth noting that the results in Table~\ref{tab:sota_compare} indicate that encoder-only fine-tuning yields higher performance than decoder-only fine-tuning for the Whisper-small model. This further supports the above conclusion by demonstrating that the performance gap cannot be attributed to Whisper models having fully mastered acoustic modeling on code-switching data.

While the above observation holds consistently across datasets, experimental results show that fine-tuning on ASRU- or LibriSpeech-accented SEAME text leads to only a moderate performance drop on SEAME data. This asymmetry suggests that the rich and complex switching patterns in SEAME provide sufficient linguistic diversity for the model to generalize effectively, even when the acoustic characteristics like accent differ. These findings underpin the conclusion that the syntactic and lexical characteristics of the training data play a more critical role than accent in determining CS-ASR performance. Moreover, the above implies that a CS-ASR model may not be robust against various levels of language confusion, such as different frequencies of code-switching or the presence of discourse markers. Consequently, this implies that training purely on code-switching data with fixed or specific textual patterns may not provide a generalizable or scalable solution for robust CS-ASR.

\subsection{Evaluation of the proposed SECT prompt}
\input{Tables/text_gpt_judge}

\input{Tables/prompt_and_eval}
We evaluated the effectiveness of the proposed SECT-guided text generation method on 1,500 synthesized text samples in terms of WER for ASR performance, GPT-4o scores, and CMI. ASR performance was evaluated by fine-tuning a Whisper-small model on \textcolor{black}{1500 synthesized speech-text pairs} before test, \textcolor{black}{where TTS is done through the CosyVoice 2 model fine-tuned on the ASRU data}. GPT-4o scores were obtained via an LLM-as-a-judge protocol, used together with CMI and \textcolor{black}{human evaluation} to assess the naturalness, grammatical correctness, and semantic preservation of the generated text samples. \textcolor{black}{PPL were computed by a Transformer LM trained on the ASRU training set, which reflects the average uncertainty in next-token prediction.}

The comparison between SECT-guided prompts and other prompting strategies is presented in Table~\ref{tab:gpt_score}. and the prompting strategies and LLM-as-a-judge evaluation are introduced in Table~\ref{tab:prompt_comparison}. The results demonstrate that the SECT-guided prompting strategy achieves the best performance in terms of WER and GPT-4o scores, and yields CMI \textcolor{black}{and PPL closest to that of the real ASRU code-switching text among all evaluated text generation strategies.}

\textcolor{black}{The PPL presented in Table~\ref{tab:gpt_score} indicates that our approach produces text that more closely aligns with the linguistic distribution of target code-switched speech. Given that the PPL values for both the test set and the generated text are significantly higher than that of the training data (4.3), we further verified this trend using a 3-gram word-level LM. The 3-gram LM achieves a PPL of 16.2 on the training data compared to 324.4 on the test data. This significant discrepancy is consistent with observations from the Transformer-based LM. These suggest that code-switching data inherently exhibits higher linguistic complexity and language confusion than monolingual data. The significantly high PPL of the generated text can likely be attributed to the presence of out-of-vocabulary (OOV) words and characters.}

\textcolor{black}{We also compared the LLM-as-a-judge with human evaluation in Table~\ref{tab:gpt_score}. Results show that the LLM-as-a-judge tends to overestimate the quality of text generated by LLM. In addition, human evaluation indicates that LLM-generated text does not yet fully replicate the linguistic distribution of spontaneous bilingual speech. Despite this gap, the proposed SECT method exhibits the highest score in the human evaluation, being consistent with LLM-as-a-judge. This demonstrates the effectiveness of the proposed SECT prompting strategy in code-switching text generation.}

Moreover, we conducted ablation studies by comparing three prompting strategies: the baseline prompt, the SECT prompt without in-context learning, and the SECT prompt with in-context learning. Results confirm the effectiveness of the SECT-based prompting strategy, since it consistently outperforms the baseline regardless of whether in-context examples are included. Although incorporating in-context learning examples does not improve the GPT-4o scores, it leads to lower WER \textcolor{black}{as well as CMI and PPL values that more closely approximate} the real ASRU data. This implies that while ECT provides beneficial structural constraints for generating linguistically valid code-switching text, it may also impose rigid switching patterns. Incorporating in-context examples helps mitigate such rigidity, leading to more natural code-switching behavior.

\subsection{TTS with text generated via SECT-prompted LLM}

\input{Tables/syn_asru_text}

We then explored incorporating the SECT-prompted LLM into TTS-based data augmentation for code-switching ASR. Specifically, code-switching texts generated by the SECT-prompted LLM were used as the target text to a TTS model in order to synthesize code-switching speech samples. Speech samples from the ASRU training set were used as reference prompts. The resulting synthetic speech–text pairs were then used for data augmentation in ASR training. 

This SECT-based augmentation strategy is compared with the TTS-based data augmentation using only original text data as the target text introduced in Section~\ref{sec:exp_tts_aug}, and the results are presented in Table~\ref{tab:syn_asru_text}. The two augmentation strategies differ in the source of their target texts. The TTS-based method draws target texts from the intra-sentence ASRU training set, while the SECT-based method obtains code-switching target texts by translating Mandarin-only ASRU training texts via an LLM guided by SECT prompts.

Results in Tables~\ref{tab:syn_data_aug} and \ref{tab:syn_asru_text} show that augmenting the training data with synthesized speech based on existing text patterns leads to degraded ASR performance on the ASRU dataset. This suggests that simply reusing the original training text during TTS fails to introduce sufficient linguistic diversity, limiting the benefit of data augmentation. In contrast, LLM-based text generation introduces greater variety into the synthesized speech–text pairs. Therefore, it exhibits higher performance than fine-tuning on the original dataset. Results also indicate that additionally including pure Mandarin data in fine-tuning degrades the performance. This finding further supports the effectiveness of the proposed SECT-guided text generation method using LLMs for code-switching data augmentation. 

\textcolor{black}{The results in Table \ref{tab:syn_asru_text} also demonstrate that the proposed SECT-based text generation method effectively augments the training of the Conformer-L model, yielding a 13.6\% relative improvement. Regarding the Whisper models, the proposed method provides consistent performance gains for the small and medium sizes. However, a performance degradation is observed when fine-tuning Whisper-large-v3. This implies that for models with significantly high parameter counts, the specific distribution of the augmented data may induce overfitting. This consequently compromises the generalization capability of the model.}

We also investigated the impact of varying the amount of synthesized speech–text pairs on CS-ASR performance. Results show that increasing the amount of pure Mandarin data during Whisper fine-tuning leads to performance degradation. This finding is consistent with the results presented in Table~\ref{tab:syn_data_aug}, where data augmentation \textcolor{black}{with limited code-switching patterns results in lower performance on ASRU data.} 

In contrast, incorporating synthesized code-switching speech–text pairs into fine-tuning yields moderate improvements. 
\textcolor{black}{With a fixed baseline of 193 hours of real speech, we evaluated three data augmentation scales to investigate the relationship between real and synthetic data. In the low augmentation setting using 100 hours of synthetic speech, the model benefits moderately from the introduction of synthetic code-switching patterns. Increasing the synthetic data to 247 hours, a balanced augmentation, exhibits the highest performance. This is attributed to doubling the exposure to varied syntactic code-switching patterns while maintaining a strong anchor in real acoustic distributions. Model performance degrades when the volume of synthetic data is further increased to 509 hours, a state of synthetic over-representation. This suggests that the model starts to overfit to the specific acoustic artifacts and prosodic regularities of the TTS model. The resulting overfitting leads to a distribution shift, consequently compromising generalization on real-world human speech. Furthermore, this trend implies that the synthesized text data still falls short of real-world text in terms of naturalness and linguistic richness.}

In addition to the high-performance DeepSeek-R1 API, we explored the use of Qwen-chat-32B, an open-sourced LLM. However, Qwen-chat-32B exhibits significantly lower performance than the DeepSeek-R1 API in terms of the quality of the generated code-switching text. With the same amount of Mandarin input sentences, the Qwen-chat-32B model successfully generates code-switching sentences corresponding to 352 hours of synthesized speech data, significantly less than 509 hours of speech data synthesized with texts generated by the DeepSeek-R1 API. This observation is reasonable, given that the DeepSeek-R1 API model has a substantially larger parameter count compared to Qwen-chat-32B. It also suggests that API-accessible models remain a strong choice for data generation, even in specialized scenarios such as code-switching.

\input{Tables/case_study}
\subsection{\textcolor{black}{Case study for error analysis}}
\textcolor{black}{Existing research has explored how interjections, such as lah, ah, and lor in SEAME, impact the CS-ASR performance~\cite{align}. In our study, the CS-ASR performance can reach 6.6\% MER for a fine-tuned Whisper-larget-v3 model, while the performance on two SEAME test sets remains high, being 10.3\% and 14.9\%, respectively. As models have shown promising performance on the ASRU data, we thus conducted a case study with representative examples as presented in Table~\ref{tab:case_study}.} 

\textcolor{black}{One common error involves phonetic confusion, where the English word, pumping, is incorrectly decoded as the Mandarin homophone. This is directly attributed to language confusion, as the decoder fails to leverage sufficient linguistic context to distinguish between the two languages at the switch point. Additionally, interjections remain a challenge for all models, which is consistent with findings in \cite{align}. The examples also show that Whisper-large-v3 often exhibits a monolingual bias with a zero-shot setup, where it fails to switch and stays in a single language.}

\section{\textcolor{black}{Discussion}}
\subsection{\textcolor{black}{Dynamic nature of code-switching}}
\textcolor{black}{The dynamic nature of code-switching results in a combinatorial complexity. Unlike monolingual speech, code-switching does not adhere to a fixed vocabulary or a single grammar~\cite{ect, muysken2000bilingual}. Instead, it represents vast combinations of two or more languages. Treating this phenomenon as a third language ignores the nature of how speakers transition between languages. These make it inapplicable to synthesize code-switching text that aligns with an unknown target textual pattern.}

\textcolor{black}{Supported by established literature~\cite{yang2023adapting, bi_encoder, yan2022joint, lae, wang2025camel}, our research suggests that utilizing monolingual data from the constituent languages significantly enhances CS-ASR performance. This trend remains consistent across different scenarios, such as the SEAME and ASRU datasets. These findings suggest that models benefit more from understanding the interaction between existing linguistic systems than from attempting to learn a new linguistic pattern from scratch.}

\subsection{\textcolor{black}{Limitations of LLM-based TTS}}
\textcolor{black}{In this study, we merely focused on fine-tuning the LLM module, which may result in limitations. In the CosyVoice 2 architecture, the LLM processes concatenated text and speech embeddings to model the conditional probability of output speech tokens. Restricting adaptation to the LLM module may lead to a cross-modal representation mismatch between the updated linguistic features and the fixed speech embeddings. This can thus limit the capability of the TTS model to achieve optimal acoustic naturalness and prosodic flow in the synthesized speech, particularly at code-switching boundaries.}

\section{Conclusion}

\textcolor{black}{In this paper, we investigated CS-ASR from both model- and data-centric perspectives, focusing on language confusion, accent bias, and data scarcity. Our model-centric analysis systematically reviewed language-aware algorithms, including language-specific processing and multi-task optimization. The evaluation indicates that these algorithms effectively mitigate language confusion. However, their performance is partially constrained by high accent bias, where language-discriminative information is inherently weakened. The data-centric analysis demonstrates that the optimal data augmentation strategy is highly dependent on the code-switching characteristics of the target domain. In high-frequency scenarios, textual patterns are naturally diverse. Therefore, the primary challenge is achieving acoustic robustness against rapid phonetic transitions. Standard augmentation methods are sufficient in these cases. Conversely, low-frequency scenarios exhibit an acoustic profile clearly biased toward the matrix language while textual variety remains inherently constrained. Consequently, the priority shifts toward increasing textual and structural diversity.}

\bibliographystyle{IEEEtran}
\bibliography{reference}

\end{document}

%% file: Tables/dataset.tex
\begin{table}[t]
\centering
\caption{Details of datasets used in terms of division and durations}
\label{tab:data}
\renewcommand{\arraystretch}{1}
\setlength{\tabcolsep}{3mm}{
\begin{tabular}{c|c|c}
\toprule
\textbf{Corpus}                 & \textbf{Subset}  & \textbf{Duration (hours)} \\ \midrule
\multirow{3}{*}{LibriSpeech~\cite{librispeech}}    & train-clean-100         & 100.6        \\ 
                                & train-clean-360         & 363.6        \\ 
                                & dev-clean     & 5.4        \\
                                & test-clean    & 5.4         \\ \midrule
\multirow{3}{*}{ASRU Mandarin~\cite{shi2020asru}}  & train   & 444.8        \\ 
                                & dev     & 29.3            \\ 
                                & test     & 25.9             \\ \midrule
\multirow{3}{*}{ASRU~\cite{shi2020asru}}           & train    & 193.0        \\ 
                                & dev      & 21.3             \\
                                & test     & 20.4             \\ \midrule
\multirow{4}{*}{SEAME~\cite{seame}} & train  & 96.6             \\ 
                       & valid    & 4.9              \\ 
                       & $\text{dev}_{\texttt{man}}$  & 7.5 \\ 
                       & $\text{dev}_{\texttt{sge}}$  & 3.9 \\ \bottomrule
\end{tabular}}
\end{table}

%% file: Tables/language_ratio.tex
\begin{table}[t]
\centering
\caption{Utterance-level duration ratios, dataset-level token distribution (English words and Mandarin characters) ratios, and CMI of the ASRU and SEAME test sets. CMI values in brackets exclude monolingual utterances}
\label{tab:seame}
\renewcommand{\arraystretch}{1.1}
\setlength{\tabcolsep}{1.5mm}{
\begin{tabular}{c|ccc|cc|c}
\toprule
\multirow{2}{*}{\textbf{Subset}}  & \multicolumn{3}{c|}{\textbf{Duration ratio (\%)}} & \multicolumn{2}{c|}{\textbf{Token ratio (\%)}} & \multirow{2}{*}{\textbf{CMI}} \\ 
          & Man & Eng & CS & Man & Eng & \\ \midrule
ASRU test                               & 0   & 0  & 100 & 89 & 11 & 11.1\\ \midrule
$\text{SEAME dev}_{\texttt{man}}$       & 14  & 7  & 79  & 74 & 26 & 16.2 (23.9)\\ 
$\text{SEAME dev}_{\texttt{sge}}$       & 6   & 41 & 53  & 37 & 63 & 12.3 (29.8)\\ \bottomrule
\end{tabular}}
\end{table}

%% file: Tables/overall.tex
\begin{table}[t]
\centering
\caption{Performance evaluation of state-of-the-art approaches on test data from the ASRU 2019 challenge and SEAME dataset in terms of MER (\%). ``$pt$", ``$adapt.$", ``$ft$" denote pre-training, adaptation tuning, and full-parameter fine-tuning, respectively. Results marked with ``*'' are taken from the original publication}
\label{tab:sota_compare}
\renewcommand{\arraystretch}{1.2}
\setlength{\tabcolsep}{1.2mm}{
\begin{tabular}{cccccc}
\toprule
\multirow{2}{*}{\textbf{Method}} & \multicolumn{1}{c}{\textbf{\#Train.}} & \multirow{2}{*}{ \textbf{$pt$} } & \multicolumn{1}{c}{\textbf{ASRU}} &  \multicolumn{2}{c}{\textbf{SEAME}} \\ \cmidrule(lr){5-6}
 & \multicolumn{1}{c}{\textbf{Params.}}&  & test$\downarrow$ & $\text{dev}_{\texttt{man}}$$\downarrow$ & $\text{dev}_{\texttt{sge}}$$\downarrow$\\ \midrule
\multicolumn{1}{l}{Conformer~\cite{conformer}}                     & 48.3 M  & \ding{55}      & 12.8  & 16.6  & 23.3 \\ 
\multicolumn{1}{l}{\textcolor{black}{+ WavLM-large~\cite{wavlm}}}   & \textcolor{black}{48.3 M}   & \textcolor{black}{\checkmark}     & \textcolor{black}{12.0}  & \textcolor{black}{15.2}  & \textcolor{black}{21.9} \\
\multicolumn{1}{l}{\textcolor{black}{+ XLS-R-0.3B~\cite{xlsr}}}     & \textcolor{black}{48.3 M}   & \textcolor{black}{\checkmark}     & \textcolor{black}{11.8}  & \textcolor{black}{15.3}  & \textcolor{black}{21.6} \\
\multicolumn{1}{l}{+ LPB~\cite{liu23_icassp}}                      & 79.9 M  & \ding{55}      & 11.8  & 16.3  & 22.9 \\ 
\multicolumn{1}{l}{+ ILB~\cite{liu2024enhancing}}                  & 79.9 M  & \ding{55}      & 11.8  & 16.4  & 23.2 \\ 
\multicolumn{1}{l}{+ LAL~\cite{align}}                             & 48.3 M  & \ding{55}      & 11.7  & 16.4  & 23.3 \\ 
\multicolumn{1}{l}{\hspace{1mm} + LLM GER~\cite{ger, align}}       & 48.3 M  & \ding{55}      & 11.0  & 15.7  & 22.0 \\ \midrule
\multicolumn{1}{l}{Transformer~\cite{transformer}}                 & 29.8 M  & \ding{55}      & 13.1  & 17.7  & 24.5 \\ 
\multicolumn{1}{l}{Branchformer~\cite{peng2022branchformer}}       & 39.0 M  & \ding{55}      & 11.9  & 16.4  & 23.2 \\ 
\multicolumn{1}{l}{E-Branchformer~\cite{kim2023branchformer}}      & 39.9 M  & \ding{55}      & 11.8  & 16.4  & 23.2 \\ 
\multicolumn{1}{l}{+ CAMEL~\cite{wang2025camel}}                        & 55.3 M  & \ding{55}      & 11.4  & 16.1  & 22.8 \\ 
\multicolumn{1}{l}{ConExtBiMamba*~\cite{mamba_in_speech}}          & 54.6 M  & \ding{55}      & 11.5  & 16.6  & 23.4 \\ \midrule
\multicolumn{1}{l}{Whisper-small~\cite{whisper}}                   & -       & \checkmark     & 24.9  & 90.8  & 69.7 \\ 
\multicolumn{1}{l}{+ $adapt.$ w/ LoRA~\cite{hu2022lora}}           & 4.4 M   & \checkmark     & 10.2  & 13.5  & 18.7 \\ 
\multicolumn{1}{l}{+ $adapt.$ w/ Bi-LoRA}                          & 4.4 M   & \checkmark     & 10.0  & 13.8  & 19.3 \\ 
\multicolumn{1}{l}{+ $ft$ encoder-only}                            & 87.0 M & \checkmark      & 9.8   & 14.6  & 19.5 \\ 
\multicolumn{1}{l}{+ $ft$ decoder-only}                            & 153.6 M & \checkmark     & 10.3  & 17.4  & 25.8 \\ 
\multicolumn{1}{l}{+ $ft$ full model}                              & 240.6 M & \checkmark     & 8.8   & 13.2  & 18.4 \\ 
\multicolumn{1}{l}{\hspace{1.2mm} + w/ SPT*~\cite{spt}}            & 240.7 M & \checkmark     & 12.8  & 13.1  & 18.7 \\ 
\multicolumn{1}{l}{\hspace{1.2mm} + w/ FA-LID-LB~\cite{align}}     & 240.6 M & \checkmark     & 8.8   & 13.3  & 18.3 \\ 
\multicolumn{1}{l}{\hspace{1.2mm} + w/ FA-LID-UB~\cite{align}}     & 240.6 M & \checkmark     & 8.7   & 13.2  & 18.3 \\ 
\multicolumn{1}{l}{\hspace{1.2mm} + w/ CTC-LID~\cite{lid_ctc}}     & 240.6 M & \checkmark     & 8.9   & 13.7  & 18.8 \\ 
\multicolumn{1}{l}{\hspace{1.2mm} + w/ LAL~\cite{align}}           & 240.6 M & \checkmark     & 8.7   & 13.3  & 18.3 \\ 
\multicolumn{1}{l}{\textcolor{black}{Whisper-medium~\cite{whisper}}}   & -       & \textcolor{black}{\checkmark}     & \textcolor{black}{12.1}  & \textcolor{black}{87.1}  & \textcolor{black}{62.6} \\ 
\multicolumn{1}{l}{\textcolor{black}{+ $ft$ full model}}               & \textcolor{black}{762.3 M}  & \textcolor{black}{\checkmark}     & \textcolor{black}{6.8}  & \textcolor{black}{11.5}  & \textcolor{black}{16.5} \\ 
\multicolumn{1}{l}{\textcolor{black}{Whisper-large-v3~\cite{whisper}}} & -       & \textcolor{black}{\checkmark}     & \textcolor{black}{10.3}  & \textcolor{black}{73.5}  & \textcolor{black}{53.3} \\ 
\multicolumn{1}{l}{\textcolor{black}{+ $ft$ full model}}               & \textcolor{black}{1541.6 M}  & \textcolor{black}{\checkmark}     & \textcolor{black}{6.6}  & \textcolor{black}{10.3}  & \textcolor{black}{14.9} \\ \bottomrule

\end{tabular}}
\end{table}

%% file: Tables/pretrain_2.tex
\begin{table}[t]
\centering
\caption{Evaluation of multilingual ASR models pre-trained on monolingual and code-switching data, with results reported before and after fine-tuning on code-switching data. Results marked with ``*'' are taken from the original publication}
\label{tab:pre-train}
\renewcommand{\arraystretch}{1.1}
\setlength{\tabcolsep}{2mm}{
\begin{tabular}{cc cc}
\toprule
\multirow{2}{*}{\textbf{Method}} & \multicolumn{2}{c}{\textbf{Monolingual}} & \textbf{CS}  \\ \cmidrule(lr){2-3} 
                                 & ASRU Man.$\downarrow$   & LS-test$\downarrow$  & ASRU test$\downarrow$\ \\ \midrule
\multicolumn{1}{l}{Conformer-L Mix~\cite{conformer}}  & 4.8  & 5.7  & 8.8     \\ \midrule
\multicolumn{1}{l}{Conformer-L Mono~\cite{conformer}} & 3.7  & 4.5  & 33.9    \\ 
\multicolumn{1}{l}{+ $ft$ ASRU}                     & 19.9 & 87.0 & 11.2    \\  
\multicolumn{1}{l}{Bi-encoder*~\cite{bi_encoder}}   & 3.3  & 9.9  & 9.8     \\ 
\multicolumn{1}{l}{+ $ft$ ASRU*~\cite{bi_encoder}}  & 5.4  & 28.0 & 9.3     \\ 
\multicolumn{1}{l}{Whisper-small~\cite{whisper}}    & 19.0 & 3.4  & 24.9   \\ 
\multicolumn{1}{l}{+ $ft$ ASRU}                     & 15.8 & 13.3 & 8.8     \\ \bottomrule

\end{tabular}}
\end{table}

%% file: Tables/syn_augment.tex
\begin{table}[t]
\centering
\caption{Comparison of fine-tuning the Whisper-small model on real, synthesized speech, and data augmented with synthesized speech. Here, the target texts and prompt samples are sourced from the original ASRU dataset. Real and synthesized data are denoted as ``Real'' and ``Syn.'', respectively. \textcolor{black}{``Speed Pert.'' is speed perturbation}}
\label{tab:syn_data_aug}
\renewcommand{\arraystretch}{1.1}
\setlength{\tabcolsep}{1.45mm}{
\begin{tabular}{cccc c}
\toprule
\multirow{2}{*}{\textbf{Method}} & \textbf{TTS}  & \textbf{ASRU} &  \multicolumn{2}{c}{\textbf{SEAME}} \\ \cmidrule(lr){4-5}
& \textbf{fine-tune} & test$\downarrow$&$\text{dev}_{\texttt{man}}$$\downarrow$ &$\text{dev}_{\texttt{sge}}$$\downarrow$  \\ \midrule
\multicolumn{1}{l}{Whisper-small}                     & -           & 24.9  & 72.3    & 53.8  \\ 
\multicolumn{1}{l}{+ $ft$ Syn. w/ Real text}          & \ding{55}   & 11.0  & 26.3    & 36.1  \\ 
\multicolumn{1}{l}{+ $ft$ Syn. w/ Real text}          & \checkmark  & 10.5  & 15.3    & 20.9  \\ \midrule 
\multicolumn{1}{l}{+ $ft$ Real data}                  & -           & 8.8   & 13.2  & 18.4  \\ 
\multicolumn{1}{l}{\hspace{1.1mm} + Syn. w/ Real text}& \checkmark  & 9.3   & 13.0    & 18.3  \\ 
\multicolumn{1}{l}{\hspace{1.1mm} + 2$\times$ Syn. w/ Real text}    & \checkmark  & 9.3   & 12.9  & 18.0  \\ 
\multicolumn{1}{l}{\hspace{1.1mm} + 3$\times$ Syn. w/ Real text}    & \checkmark  & 9.4   & 12.7  & 17.7  \\ 
\multicolumn{1}{l}{\hspace{1.1mm} \textcolor{black}{+ Speed Pert. (w/ 0.9 \& 1.1)}}    & \textcolor{black}{-}   & \textcolor{black}{9.1}  & \textcolor{black}{12.9}  & \textcolor{black}{18.3}  \\ 
\multicolumn{1}{l}{\hspace{3.4mm} \textcolor{black}{+ 3$\times$ Syn. w/ Real text}}  & \textcolor{black}{\checkmark}  & \textcolor{black}{9.1}  & \textcolor{black}{12.6}  & \textcolor{black}{17.2}  \\ \bottomrule

\end{tabular}}
\end{table}

%% file: Tables/tts_data_mismatch.tex
\begin{table}[t]
\centering
\caption{Comparison of textual and acoustic impacts on synthesized data for fine-tuning the Whisper-small model. Text and Prompt denote the source of synthesized speech and the prompt speech, respectively. TTS $ft$ denotes the data type used for fine-tuning the TTS model}
\label{tab:mismatch}
\renewcommand{\arraystretch}{1.1}
\setlength{\tabcolsep}{1.0mm}{
\begin{tabular}{cccccc c}
\toprule
\multirow{2}{*}{\textbf{Mismatch}} &\multirow{2}{*}{\textbf{Target}} & \multirow{2}{*}{\textbf{Reference}} &  \multirow{2}{*}{\textbf{TTS} $ft$} & \multicolumn{2}{c}{\textbf{SEAME}} & \textbf{ASRU}\\ \cmidrule(lr){5-6}
& & & &$\text{dev}_{\texttt{man}}$$\downarrow$ &$\text{dev}_{\texttt{sge}}$$\downarrow$ & test$\downarrow$  \\ \midrule
-                & SEAME & SEAME & SEAME        & 15.3 & 20.9 & -    \\
\textbf{Accent}  & SEAME & ASRU  & SEAME        & 16.1 & 22.1 & -    \\
\textbf{Accent}  & SEAME & LibriSpeech  & SEAME & 17.4 & 25.5 & -    \\
\textbf{Text}    & ASRU  & SEAME & SEAME        & 35.1 & 68.2 & -    \\ \midrule
-                & ASRU  & ASRU  & ASRU         & -    & -    & 10.1 \\
\textbf{Accent}  & ASRU  & SEAME & ASRU         & -    & -    & 10.9 \\
\textcolor{black}{\textbf{Accent}}  & \textcolor{black}{ASRU}  & \textcolor{black}{LibriSpeech} & \textcolor{black}{ASRU}   & -    & -    & \textcolor{black}{10.9} \\
\textbf{Text}    & SEAME & ASRU  & ASRU         & -    & -    & 16.6 \\ \bottomrule

\end{tabular}}
\end{table}

%% file: Tables/text_gpt_judge.tex
\begin{table}[t]
\centering
\caption{Comparison of code-switching text synthesis with different LLMs and prompts, performance is evaluated with code-switching samples synthesized via 1500 ASRU Mandarin text samples, where the model is \textcolor{black}{fine-tuned on 1500 synthesized speech-text pairs} and tested on the ASRU test set. The CMI of the ASRU training set is 17.0, \textcolor{black}{the PPL of the ASRU test set is 310.8}}
\label{tab:gpt_score}
\renewcommand{\arraystretch}{1.2}
\setlength{\tabcolsep}{1.3mm}{
\begin{tabular}{cccccc}
\toprule
\textbf{Prompting method}        & \textbf{ASR}  & \textbf{GPT-4o} & \textcolor{black}{\textbf{Human}} & \textbf{CMI} & \textcolor{black}{\textbf{PPL}}\\
\textbf{DeepSeek-R1 API}         & WER$\downarrow$  & score$\uparrow$ & \textcolor{black}{score$\uparrow$} & $gt$: 17.0 &\textcolor{black}{$gt$:310.8} \\ \midrule
\multicolumn{1}{l}{Baseline prompt}                                 & 22.9    & 6.8  & \textcolor{black}{3.9}    & 32.2 & \textcolor{black}{1794.5}\\ 
\multicolumn{1}{l}{EZSwitch prompt~\cite{kuwanto2024linguistics}}   & 22.2    & 6.0  & \textcolor{black}{5.0}    & 24.6 & \textcolor{black}{934.5}\\ \midrule
\multicolumn{1}{l}{SECT prompt (ours)}                              & \textbf{19.2}  & \textbf{7.0} & \textcolor{black}{\textbf{6.7}}   & \textbf{17.3}  & \textcolor{black}{\textbf{540.2}} \\ 
\multicolumn{1}{l}{\hspace{1mm} w/o \textcolor{black}{few-shot learning}}   & 19.6    & \textbf{7.0} & \textcolor{black}{5.7}   & 15.2  & \textcolor{black}{564.5}  \\ \bottomrule

\end{tabular}}
\end{table}

%% file: Tables/prompt_and_eval.tex
\begin{table}[t]
\centering
\caption{Prompts used for code-switching text generation and evaluation}
\label{tab:prompt_comparison}
\renewcommand{\arraystretch}{1.1}
\setlength{\tabcolsep}{1pt}
\begin{tabular}{p{1.5cm}|p{7cm}}
\toprule
\multicolumn{1}{c|}{\textbf{Method}} & \multicolumn{1}{c}{\textbf{Prompt}}\\
\midrule
\multicolumn{1}{c|}{Baseline}&
You are a bilingual English–Mandarin code-switching expert. Your task is to translate the given Mandarin sentence into a natural code-switching sentence.  \\
& \texttt{<Input Sentence>} \\
\midrule
\multicolumn{1}{c|}{EZSwitch} &
You are a bilingual English-Mandarin code-switching expert. Your task is to translate the given Mandarin sentence into a code-mixed sentence with Romanized English and Mandarin with specific keywords that should appear. \\
& \texttt{<Input Sentence>} \\
& Words wanted: \texttt{<List of Words>} \\
\midrule
\multicolumn{1}{c|}{SECT (ours)} &
You are a bilingual English–Mandarin code-switching expert. Your task is to translate the given Mandarin sentence into a natural code-switching sentence. Please strictly follow the rules below: 

1. Preserve the full meaning of the original sentence.

2. At most two English words or short phrases may substitute for Mandarin words in each sentence.

3. The code-switching output must be natural and fluent, consistent with the daily speech habits of Mandarin–English bilinguals.

There are three examples: 

\hspace{1.1mm} \textit{Example 1. Mandarin: ... Code-switching: ...}

\hspace{1.1mm} \textit{Example 2. Mandarin: ... Code-switching: ...}

\hspace{1.1mm} \textit{Example 3. Mandarin: ... Code-switching: ...}

Please output the code-switching sentence directly without  extra text or explanation.: \\
& \texttt{<Input Sentence>} \\
\midrule
\multicolumn{1}{c|}{GPT Eval} &
You are an expert evaluator of code-switched text quality. You will evaluate Mandarin-English code-switched sentences based on the following criteria:

1. Naturalness: How natural and fluent does the code-switched sentence sound?

2. Grammar: Is the sentence grammatically correct in both languages?

3. Semantic preservation: Does the code-switched version preserve the meaning of the original Mandarin?

4. Code-switching appropriateness: Are the language switches natural and appropriate?

Please provide a single score from 1 to 10 (where 10 is excellent and 1 is very poor). 
Return ONLY the numerical score as a single digit, nothing else. \\
& Original Mandarin: \texttt{<Original Mandarin>} \\
& Code-Switching text: \texttt{<Code-Switching text>} \\
\bottomrule
\end{tabular}
\end{table}

%% file: Tables/syn_asru_text.tex
\begin{table}[t]
\centering
\caption{Comparison between \textcolor{black}{developing ASR} models on ASRU data with different data augmentation methods. Text synthesis is performed via LLM guided by the proposed SECT prompts, and TTS is performed using target texts sourced from ``Text source'' and reference prompts sourced from the ASRU training set. Real and synthesized data are denoted as ``Real'' and ``Syn.'', respectively}
\label{tab:syn_asru_text}
\renewcommand{\arraystretch}{1.1}
\setlength{\tabcolsep}{1.5mm}{
\begin{tabular}{cccc}
\toprule

\multirow{2}{*}{\textbf{Method}}                            & \textbf{Text}         & \textbf{Data Dur.} & \textbf{ASRU}     \\
                                                            & \textbf{source}       & (hours)              & test$\downarrow$  \\ \midrule
\multicolumn{1}{l}{Whisper-small}                          & -                     & -           & 24.9    \\ 
\multicolumn{1}{l}{+ $ft$ ASRU}                            & -                     & 193         & 8.8     \\ 
\multicolumn{1}{l}{\hspace{1.2mm} + ASRU Man.}            & -                     & 193+100     & 8.9     \\ 
\multicolumn{1}{l}{\hspace{1.2mm} + ASRU Man.}            & -                     & 193+500     & 9.2     \\ 
\multicolumn{1}{l}{\hspace{1.2mm} + Syn. w/ Real text}        & ASRU                  & 193+180     & 9.3     \\ 
\multicolumn{1}{l}{\hspace{1.2mm} + Syn. text w/ DeepSeek}    & ASRU Mandarin             & 193+100     & 8.5     \\ 
\multicolumn{1}{l}{\hspace{1.2mm} + Syn. text w/ DeepSeek}    & ASRU Mandarin             & 193+247     & 8.3     \\ 
\multicolumn{1}{l}{\hspace{1.2mm} + Syn. text w/ DeepSeek}    & ASRU Mandarin             & 193+509     & 8.4      \\ 
\multicolumn{1}{l}{+ $ft$ Syn. w/ Real text}               & ASRU                  & 180         & 10.5    \\ 
\multicolumn{1}{l}{+ $ft$ Syn. text w/ Qwen}               & ASRU                  & 352         & 18.4    \\ 
\multicolumn{1}{l}{+ $ft$ Syn. text w/ DeepSeek}           & ASRU                  & 509         & 13.8    \\ \midrule

\multicolumn{1}{l}{\textcolor{black}{Conformer-L ASRU}}        & \textcolor{black}{-}   & \textcolor{black}{193}           & \textcolor{black}{11.8}    \\ 
\multicolumn{1}{l}{\textcolor{black}{+ Syn. text w/ DeepSeek}} & \textcolor{black}{ASRU Mandarin}             & \textcolor{black}{193+247}     &  \textcolor{black}{10.2}     \\ \midrule
\multicolumn{1}{l}{\textcolor{black}{Whisper-medium}}                         & \textcolor{black}{-}                     & \textcolor{black}{-}             &  \textcolor{black}{12.1}    \\ 
\multicolumn{1}{l}{\textcolor{black}{+ $ft$ ASRU}}                            & \textcolor{black}{-}                     & \textcolor{black}{193}           &  \textcolor{black}{6.8}    \\ 
\multicolumn{1}{l}{\textcolor{black}{\hspace{1.2mm} + Syn. text w/ DeepSeek}}    & \textcolor{black}{ASRU Mandarin}             & \textcolor{black}{193+247}     &  \textcolor{black}{6.6}     \\ \midrule
\multicolumn{1}{l}{\textcolor{black}{Whisper-large-v3}}                       & \textcolor{black}{-}                     & \textcolor{black}{-}             &  \textcolor{black}{10.3}    \\ 
\multicolumn{1}{l}{\textcolor{black}{+ $ft$ ASRU}}                            & \textcolor{black}{-}                     & \textcolor{black}{193}           &  \textcolor{black}{6.6}     \\ 
\multicolumn{1}{l}{\textcolor{black}{\hspace{1.2mm} + Syn. text w/ DeepSeek}}    & \textcolor{black}{ASRU Mandarin}             & \textcolor{black}{193+247}     &  \textcolor{black}{6.8}     \\ \bottomrule

\end{tabular}}
\end{table}

%% file: Tables/case_study.tex
\begin{CJK*}{UTF8}{gbsn}
\begin{table}[ht]
\centering
\caption{\textcolor{black}{A case study for error analysis of CS-ASR on SEAME dataset, with Conformer model, fine-tuned Whisper-small model, zero-shot Whisper-large-v3 model, and fine-tuned Whisper-large-v3 model}}
\label{tab:case_study}
\setlength{\tabcolsep}{0.9mm}
\begin{tabular}{l|l}
\toprule
\textbf{\textcolor{black}{Method}} & \textbf{\textcolor{black}{Output}} \\ \midrule
\multirow{3}{*}{\textcolor{black}{Conformer}}            & \textcolor{black}{not really ah sometime 我在家有自己帮兵}   \\
                                      & \textcolor{black}{relax ah mm 你会 meet friends 吗对不对}    \\          
                                      & \textcolor{black}{还有 opportunity 就呃那个呃} \\ \midrule
\multirow{3}{*}{\textcolor{black}{Whisper-small $ft$}}   & \textcolor{black}{not really lah sometimes 我在家有自己帮人做}   \\
                                      & \textcolor{black}{relax ah en 你会 meet friends 吗对不对}    \\
                                      & \textcolor{black}{还有 opportunity 就是呃去那个呃} \\ \midrule
\multirow{3}{*}{\textcolor{black}{Whisper-large-v3}}& \textcolor{black}{not really sometimes i will just pump} \\
                                      & \textcolor{black}{relax you will meet friends right}                       \\
                                      & \textcolor{black}{and the opportunity to go to}  \\ \midrule
\multirow{3}{*}{\textcolor{black}{Whisper-large-v3 $ft$}}& \textcolor{black}{not really lah sometimes 我在家有自己帮忙}  \\
                                      & \textcolor{black}{relax ah en 你 会 meet friends mah 对不对}   \\
                                      & \textcolor{black}{还有 opportunity 就是呃去那个呃}  \\ \midrule
\multirow{3}{*}{\textcolor{black}{Groundtruth}}          & \textcolor{black}{not really lah sometime 我在家里有自己 pumping}   \\
                                      & \textcolor{black}{relax ah hum 因为 meet friends 嘛对不对}             \\ 
                                      & \textcolor{black}{还有 opportunity 就是嗯去那个嗯} \\ \bottomrule
\end{tabular}
\end{table}
\end{CJK*}